\documentclass[a4paper,fleqn,usenatbib]{mnras}

\usepackage{newtxtext,newtxmath}
\usepackage{subfig}
\usepackage[T1]{fontenc}
\usepackage{ae,aecompl}

\usepackage{graphicx}	
\usepackage{amsmath}	
\usepackage{amssymb}	

\def\ut#1{\mathop{\vtop{\ialign{##\crcr
     $\hfil\displaystyle{#1}\hfil$\crcr\noalign
     {\kern1pt\nointerlineskip}\hbox{$\hfil\sim\hfil$}\crcr
     \noalign{\kern1pt}}}}}

\def\undersymbol#1#2{\mathop{\vtop{\ialign{##\crcr
     $\hfil\displaystyle{#2}\hfil$\crcr\noalign
     {\kern1pt\nointerlineskip}\hbox{$\hfil#1\hfil$}\crcr
     \noalign{\kern1pt}}}}}
\def\arcsec{^{\prime\prime}}
\def\arcmin{^{\prime}}
\def\degr{^\circ}
\def\hour{^{\rm h}}
\def\minute{^{\rm m}}
\def\second{^{\rm s}}

\title[HP Cet and Swift J0820.6-2805 in x-rays]{HP Cet and Swift J0820.6-2805: two candidate intermediate polars observed by {\it XMM}-Newton}

\author[A.A. Nucita, F. De Paolis, F. Strafella, D. Licchelli]{
A.A. Nucita$^{1,2}$\thanks{E-mail: nucita@le.infn.it},
F. De Paolis $^{1,2}$,
F. Strafella $^{1,2}$,
D. Licchelli $^{3,4}$
\\
$^{1}$ Department of Mathematics and Physics {\it ``E. De Giorgi''} , University of Salento, Via per Arnesano, CP-I93, I-73100, Lecce, Italy\\
$^{2}$  INFN, Sezione di Lecce, Via per Arnesano, CP-193, I-73100, Lecce, Italy\\
$^{3}$ R.P. Feynman Observatory, I-73034, Gagliano del Capo, Lecce, Italy  \\ 
$^{4}$ CBA, Center for Backyard Astrophysics - I-73034, Gagliano del Capo, Lecce, Italy\\
}

\date{Accepted XXX. Received YYY; in original form ZZZ}

\pubyear{2017}

\begin{document}
\label{firstpage}
\pagerange{\pageref{firstpage}--\pageref{lastpage}}
\maketitle

\begin{abstract}
We report on the $XMM$-Newton observation of HP Cet and Swift J0820.6-2805, two X-ray photon sources that are candidates to be members of the intermediate polar class of cataclysmic variables. If the historical optical light curve of HP Cet shows a periodic feature at $\simeq 96$ minutes, a clear identification of such a signature in the high energy band (apart for a variability on a time scale of $\simeq 8$ minutes as detected by the ROSAT satellite) is lacking. By using $XMM$-Newton archive data, we clearly identify a feature (at $\simeq 88$ minutes) which is marginally consistent with one of the binary system orbital periods  reported in {the} literature. We also found a signature of a periodic features on the time scale of $\simeq 5.6$ minutes. In the case of Swift J0820.6-2805, the intermediate polar nature {was previously unclear} and the orbital and the white dwarf spin periods were unknown. Here, the 0.3-10 keV data undoubtedly reveal an orbital period and a white dwarf spin of $\simeq 87.5$ minutes and $\simeq 27.9$ minutes, respectively.
{The spectral analysis showed that both HP Cet and Swift J0820.6-280 are members of the under-luminous IP subclass since 
their luminosity in the $0.3-10$ keV band is estimated to be
$\simeq 5\times 10^{30}$ erg s$^{-1}$ and $\simeq 3.8\times 10^{29}$ erg s$^{-1}$, respectively.
}

\end{abstract}

\begin{keywords}
(stars:) novae, cataclysmic variables; X-rays: binaries; X-rays: individual: HP Cet; X-rays: individual: Swift J0820.6-2805; (stars:) white dwarfs
\end{keywords}



\section{Introduction}

A binary system made of a white dwarf (WD) primary accreting material from a donor star constitutes what is commonly known as a cataclysmic variable, hereinafter CV (see, e.g. \citealt{Kuulkers2006} for a review). 

CVs {come} in three different flavours depending on the details of the accretion mechanism and the strength of the magnetic field. In particular, non magnetic systems (with a field $\ut< 0.1$ MG, see, e.g. \citealt{vanteeseling1996,nucita2009,hoard2010,nucita20092,nucita2011,balman2011,nucita2014,mukai2017}) constitute the {\it dwarf novae} class. Conversely, large magnetic field objects {(with magnetic field strength exceeding $10$ MG, see, e.g., \citealt{ramsay2004, szkody2004})} are known as {\it polars}. CVs characterized by a magnetic field value between the previous two extrema specify the subset of {\it intermediate polars} or IPs. 

In intermediate polars (see, e.g., \citealt{mukai2020} for a recent analysis of a sample of newly confirmed IPs) the accreted material suffers of a shock wave close to the WD surface and, as a consequence, it releases hard $X$-ray photons that, once detected, can be used to study in detail the accretion process. 
IP candidates are found by using optical and/or $X$-ray observations by means of the identification of emission lines and the characteristics of the high energy spectrum.

Often the high energy signal is modulated (\citealt{parker2005}) on the WD spin $P_{spin}$, the orbital period $P_{orb}$ as well as on the synodic period $P_{syn}$ (i.e.  $P_{syn}^{-1}=P_{spin}^{-1}-P_{orb}^{-1}$) {due to the presence of reprocessing sites}. Hence, the existence of multiple periodic {components} is a signature that IPs are intrinsically different from the more strongly magnetic polars in which the WD spin and the orbital period are synchronized. 

In fact, as pointed out by \citet{warnerbook}, when one considers that the $X$-ray production site is characterized by a variable geometry as seen by the observer, multiple orbital sidebands are found and naturally expected. Hence, the power spectrum is expected to host features at the additional peculiar frequencies $P_{spin}^{-1} \pm P_{orb}^{-1}$ and $P_{spin}^{-1}\pm 2P_{orb}^{-1}$ due to amplitude modulation at $P_{orb}$ and $2P_{orb}$.  

As noted by \citet{nucita2019} (but see also \citealt{mukai2020}) in the case of the IP DW Cnc, detecting (in the Fourier power spectrum of  $X$-ray data) signatures of the {orbital and spin periods, together with the associated multiple sidebands} is a powerful method for classifying a cataclysmic variable as a member of the intermediate polar class. Therefore, dedicated $X$-ray campaigns with {sufficient sensitivity} are crucial for a correct classification of such objects.

{IPs are expected to be quite common in the Galaxy and are thought to significantly  contribute  to the overall $X$-ray ridge background. In fact, it is known that much of the Galactic ridge $X$-ray  background (a diffuse emission which is  observed to be concentrated around the Galactic plane, see \cite{worrall1982}), is still without an {exhaustive explanation}. The issue was {partly} solved by using Chandra $X$-ray observations \citep{revnivtsev2009} which showed that $\simeq 80\%$ of the signal is resolved in many faint point sources, including IPs which dominate \citep{warwick2014} for energy above $\simeq 10$ keV. However, only about {\it sixty} sources have been classified as  IPs so far (see e.g. the most updated IP catalogue -{\it IPhome}- available at \url{https://asd.gsfc.nasa.gov/Koji.Mukai/iphome/iphome.html}). Therefore, dedicated survey and follow-up observations, with the aim to discover and study this population of objects, would be required.  } 

Here we report on the timing analysis of $X$-ray data from two CVs, HP Cet and Swift J0820.6-2805. {These sources are IP candidates and are present in the IPhome list where their IP nature is reported as {\it doubtful} and {\it possible}, respectively}. Furthermore, the $XMM$-Newton observations for HP Cet (ID 673140201) and  Swift J0820.6-2805 (ID 0801780401) were not previously analyzed. In particular for Swift J0820.6-2805, a confirmation of the IP nature and the identification of the orbital and spin periods are still lacking.

For HP Cet, our analysis of the 0.3-10 keV band data, as well as the UV light curve, confirms the existence of what we tentatively identify as the orbital modulation at $\simeq 88$ minutes (slightly different from the value of $\simeq 96$ {minutes} reported by \cite{southworth2006}. We further detect a periodic feature at $\simeq 5.6$ minutes that is close to variability of about $8$ minutes \citep{szkody2002}, estimated on the basis of ROSAT satellite data.

In the case of the IP candidate Swift J0820.6-2805, we showed that the high energy light curve is characterized by a modulation on what we identify to be the orbital period of {$\simeq 87.5$ minutes}. Searching for the typical sidebands expected for IPs resulted in the clear identification of the spin period at {$\simeq 27.9$ minutes}. Thus, the period search allowed us to confirm the IP nature of the source.  

\section{The {\it XMM}-Newton view of HP Cet and Swift J0820.6-2805}

\subsection{Data reduction}
\label{datareduction}

HP Cet (with J2000 coordinates ${\rm RA=02\hour 33\minute 22.61\second}$  and ${\rm DEC = +00\degr 50\arcmin 59.5\arcsec}$) was observed by the $XMM$-Newton satellite in 2012 (Observation ID 0673140201) for $\simeq 22.5$ ks. The observation started (ended) on $2012/01/15$ at $21:24:20$ ($03:39:33$) UT. The source was pointed by the EPIC pn and MOS cameras operating in full frame mode and with medium filter as well as by the Optical Monitor (OM) in image and fast mode. The adopted filter is the UVM2 which offers a bandpass centred at $\simeq 231$ nm. 

Swift J0820.6-2805 (also known as 1RXS J082033.6-280457 and PBC J0820.4-2801 with J2000 coordinates ${\rm RA=08\hour 20\minute 34.11\second}$  and ${\rm DEC = -28\degr 04\arcmin
58.8\arcsec}$) was discovered by means of the Swift survey \citep{baumgartner2013}. A bright counterpart was identified in the UV \citep{cusumano} and recognized as a possible IP by \citet{parisi2014} via the identification of prominent lines of $H$, $He$I, and $He$II.
The source was observed by the $XMM$-Newton satellite as a part of a larger follow-up program. The target was observed in 2018  (ID 801780401) by the EPIC cameras operating in small window mode and thin filter. The OM observed the target in image and fast mode with the adoption of the B filter centred at 450 nm. The observation lasted for $\simeq 37$ ks. 

The EPIC raw data files (ODFs) were processed using the $XMM$-Science Analysis System (SAS version 17.0.0) using the latest calibration constituent files (CCFs). We obtained the calibrated event list files for the three EPIC cameras by running the SAS tasks {\it emchain} and {\it epchain}. We corrected the event files for the Solar System barycenter (via the {\it barycen} SAS tool) in order to convert the photon arrival times  from spacecraft time to the barycentric dynamical time.

We then searched for segments of each observation affected by soft proton flares in order to determine a list of good time intervals (GTIs) to be applied in the following spectral and timing analysis. {Note, however, that while the observation of Swift J0820.6-2805 were not disturbed by flares \footnote{The net count rate remains in all the cameras always below $0.4$ count s$^{-1}$ for photons with energy larger than 10 keV.}, HP Cet was affected by strong flares for $\simeq 45\%$ of the observation duration. In the latter case, we decided to be as restrictive as possible and concentrate only on the first $\simeq 9$ ks }which were not affected by flares \footnote{
In order to exploit the full observation, we attempted a procedure avoiding the application of GTIs. Indeed, since any background extraction region is affected by soft proton flares similarly to what happens for the source, after a proper subtraction (bin by bin) of the synchronized light curves one is left with the full source light curves. We note that this approach introduces too many spurious features that are {\it unstable} versus a time resolved periodogram. Thus,  this possibility is not explored further in the rest of the paper.}. Obviously, this strongly reduces our capability to test periods larger than $\simeq 50$ minutes\footnote{The upper limit on the period that can be tested derives from our  request to have at least three full cycles in the timeseries.}. 

\begin{figure*}
  \centering
  \subfloat[]{\includegraphics[width=0.5\textwidth]{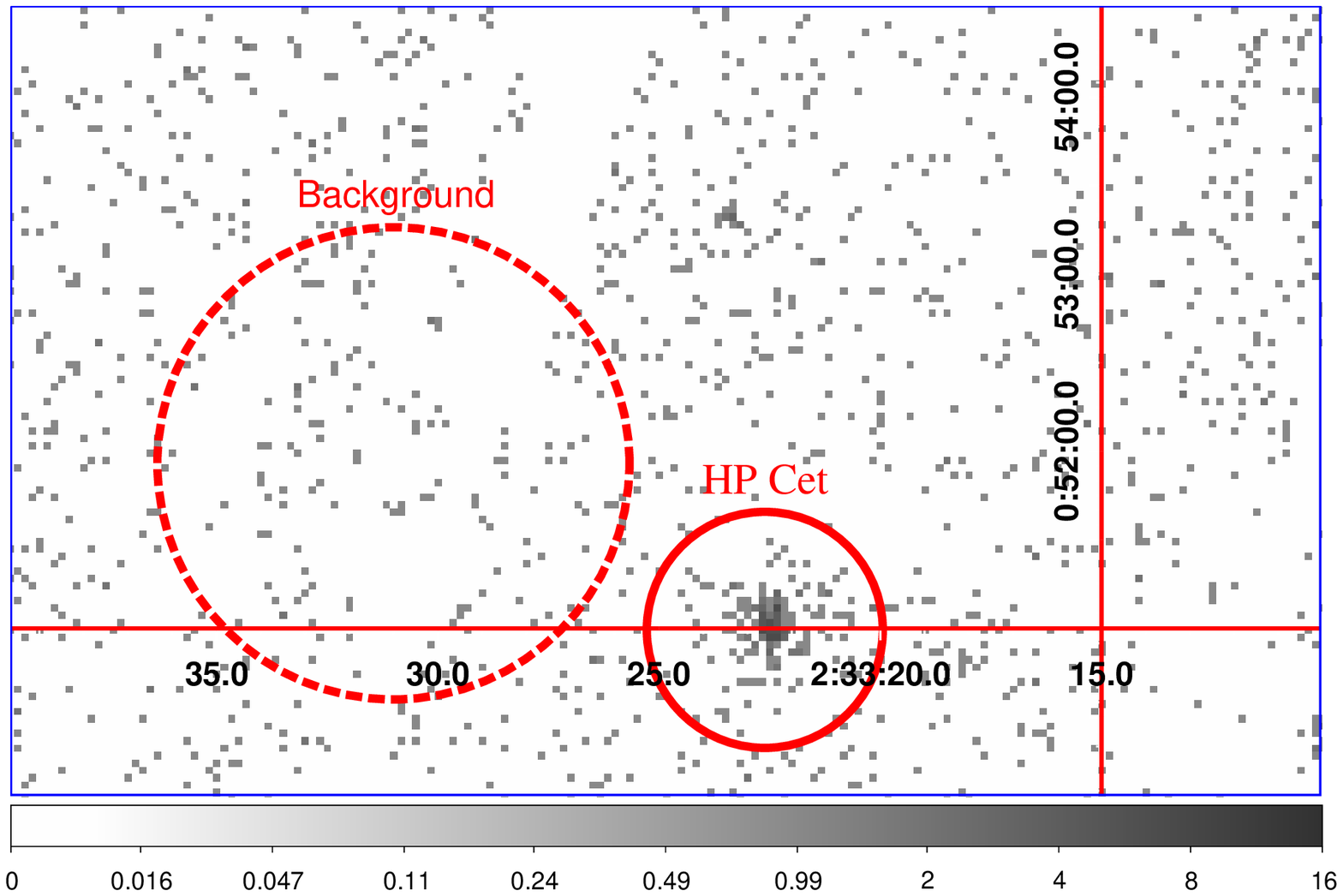}\label{fig1a}}
  \hfill
  \subfloat[]{\includegraphics[width=0.5\textwidth]{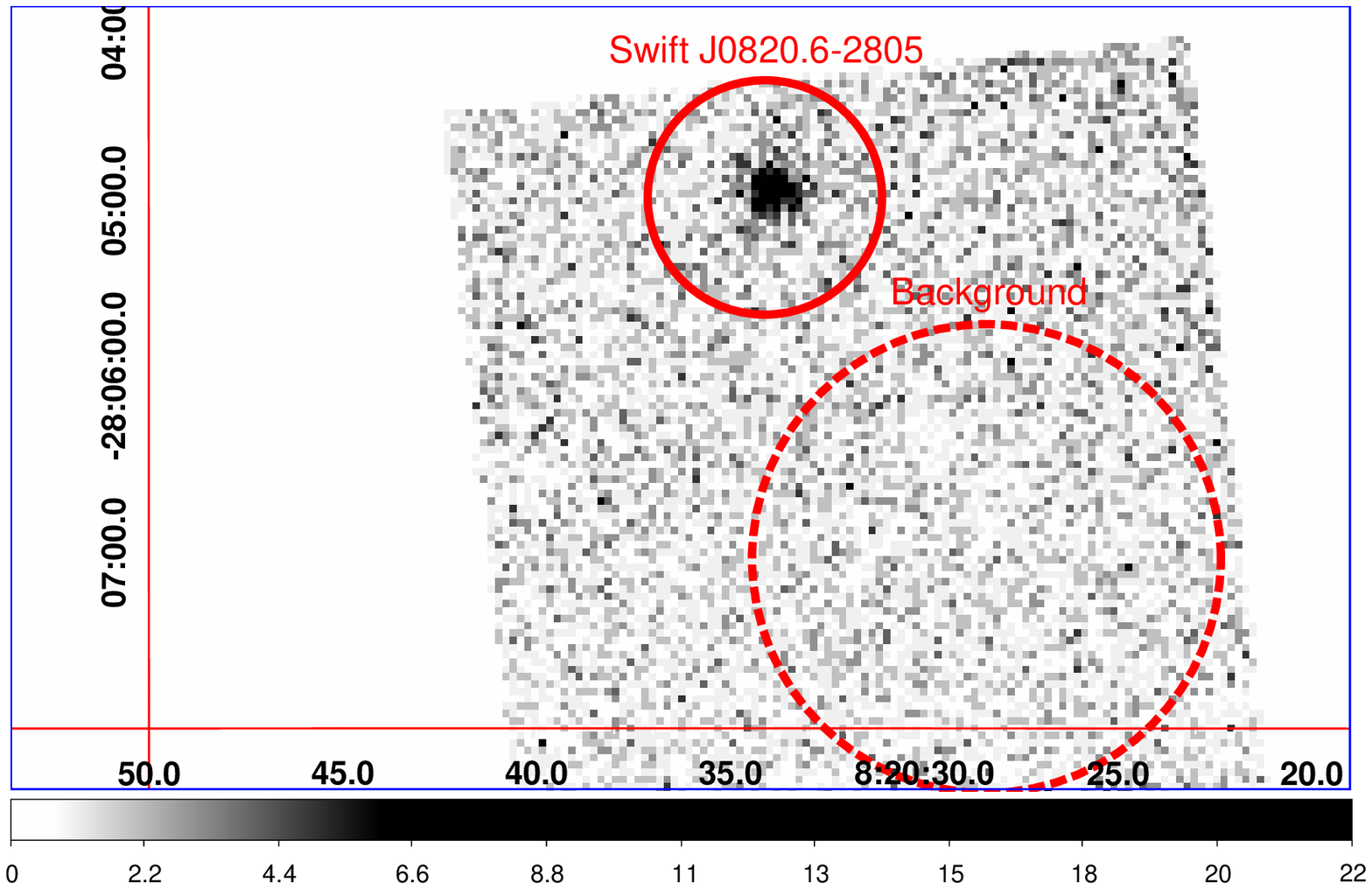}\label{fig1b}}
  \caption{{Images in the $0.3-10$ keV band were extracted as described in the text. In Figure, we give a zoom around the nominal coordinates of the target: HP Cet is on the left panel (MOS 1 camera), while Swift J0820.6-2805 is shown on the pn camera (right panel).}}
  \label{fig_ds9}
 \end{figure*}

GTIs were also considered for the spectral analysis of the source. We also generated $0.3-10$ keV images for inspecting purposes (see, e.g., Figures \ref{fig1a} and \ref{fig1b} where we give a zoom around the nominal target coordinates). 
In each case, the source (plus background) count rate was extracted in the soft ($0.3-2$ keV), hard ($2-10$ keV) and full ($0.3-10$ keV) bands by adopting a circular region with radius of $40\arcsec$ which guarantees the possibility to collect $\simeq 88\%$ of the total energy. The source extraction region was positioned on the nominal target coordinates while the background photons are extracted in circular regions (with radius of $\simeq 80\arcsec$) placed on the same chip but far from any other visible source.

\begin{figure*}
  \centering
  \subfloat[]{\includegraphics[width=0.5\textwidth]{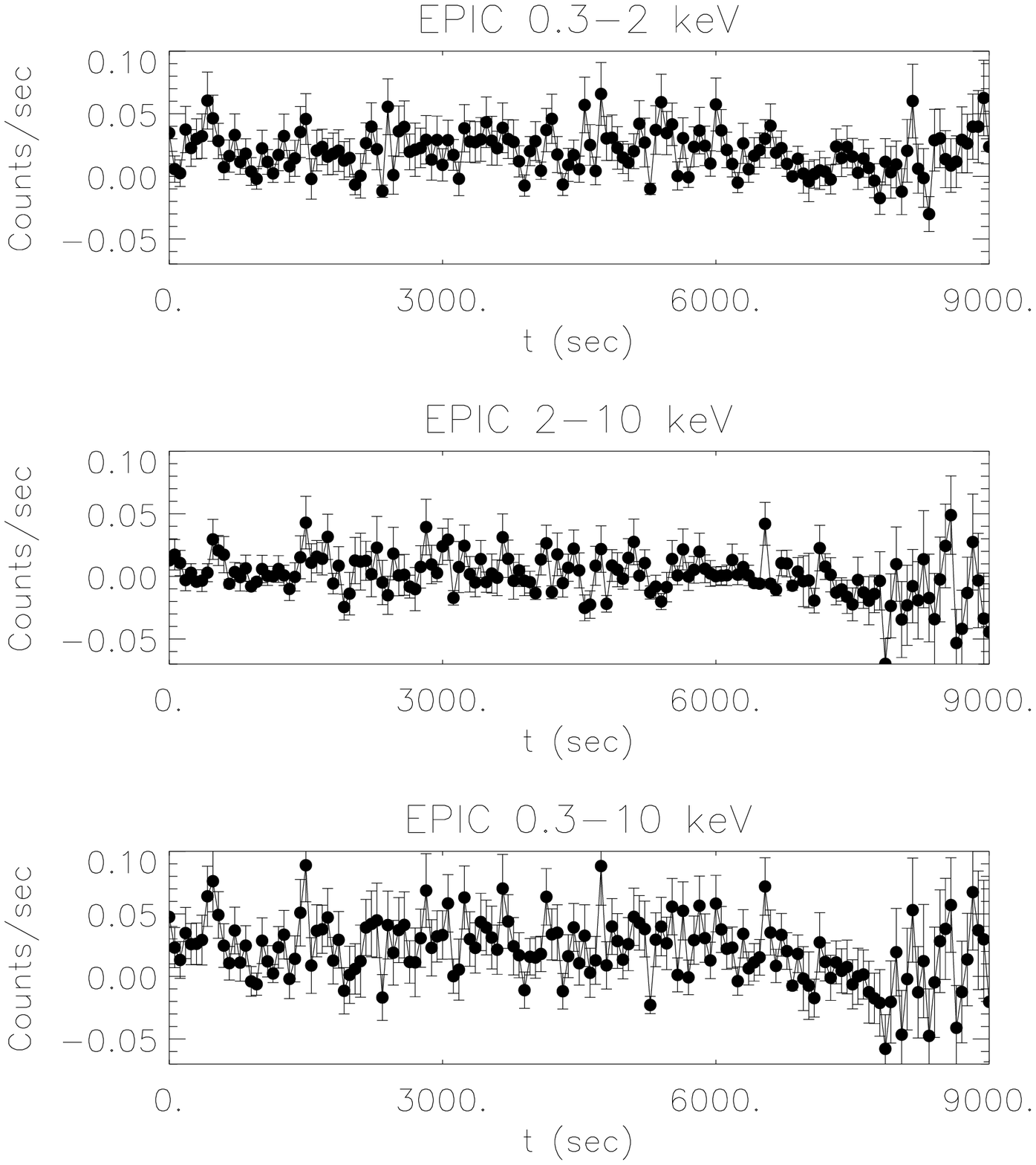}\label{fig2a}}
  \hfill
  \subfloat[]{\includegraphics[width=0.5\textwidth]{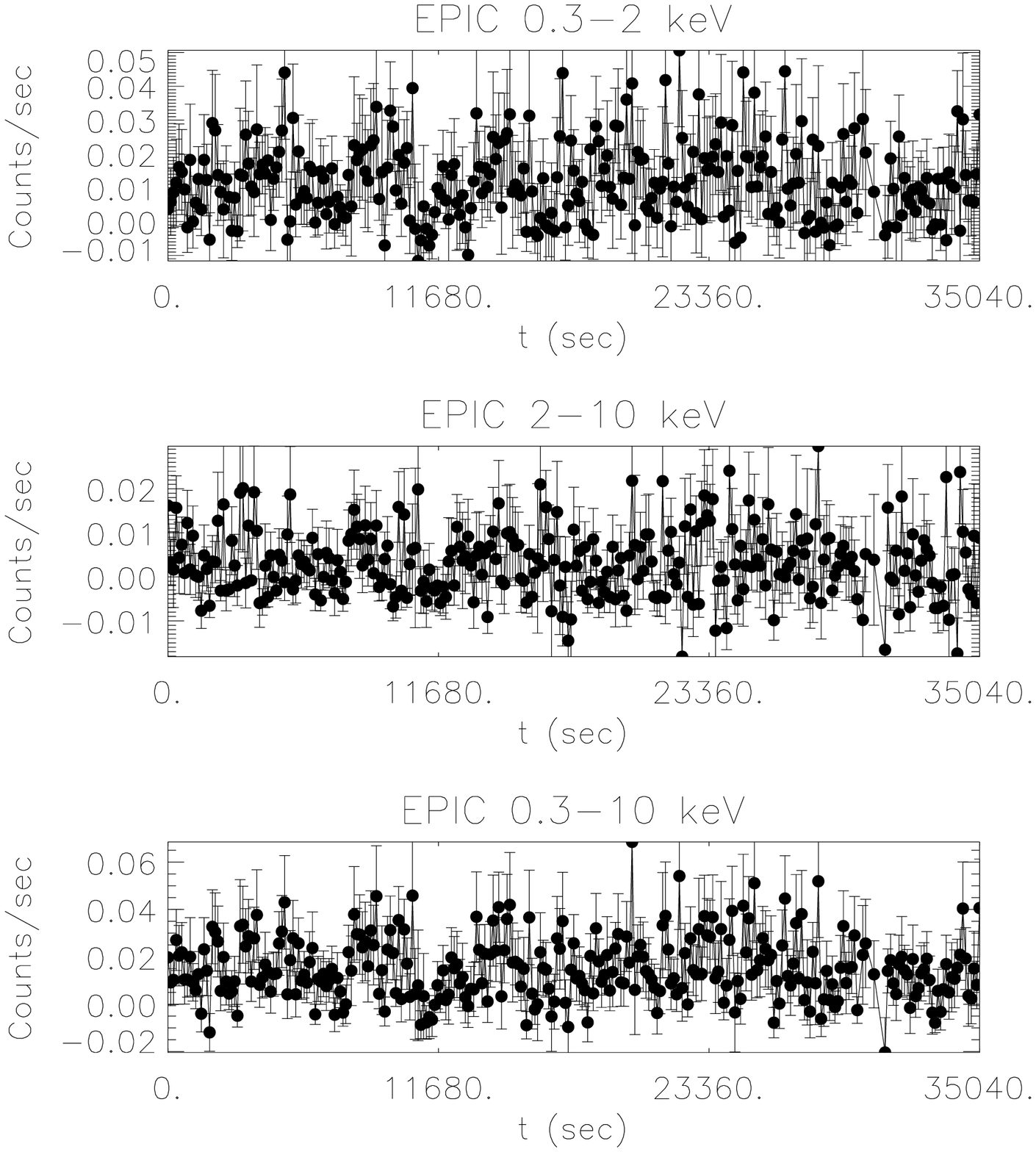}\label{fig2b}}
  \caption{Left panel: the HP Cet Epic (background subtracted and synchronized) light curves in the 0.3-2 keV, 2-10 keV and 0.3-10 keV bands, respectively. 
  Each light curve has a bin size of {60 seconds and starts at   ${\rm MJD = 55941.9103}$ days.} Right panel: the light curve extracted for Swift J0820.6-2805 starting at {${\rm MJD = 58226.3444}$ days. In this case the light curve bin size was set to $120$ seconds for graphical purposes. Each light curve is the average of the MOS 1, MOS 2 and pn (background subtracted and synchronized) light curves. See text for details.} }
  \label{fig_lc}
\end{figure*}

{
For each band and for each instrument, we first extracted the source time-series, we searched for the maximum overlapping time interval and then flagged the common start and stop times. We then produced  synchronized light curves with bin size of $10$, $60$ and $120$ seconds and repeated the whole  procedure for the background. Finally, we used the {\it epiclccorr} task to get the final MOS 1, MOS 2 and pn source (background subtracted and synchronized) time series that were then averaged bin-by-bin. The light curves were then scaled in time in order to start from $0$.  In Figures \ref{fig2a} and \ref{fig2b} we give the soft, hard and full $X$-ray light curves for the sources of interest, with the medium and hard light curves appearing more noisy.
}

The OM UVM2 (for HP Cet) and B (for Swift J0820.6-2805) data where extracted by using the standard SAS task {\it omfchain}. The bin size of each light curve\footnote{The zero point magnitudes for the OM UVM2 and B filter are 15.7724 and 19.2661, respectively \citep{xrp}.} was set to $10$ seconds. 
\label{timing}

Finally, we extracted the spectrum for the source and the background by using the same regions described above and requiring to have at  least 25 counts per energy bin. The source and background spectra as well as the associated response matrices and ancillary files were then used within the XSPEC software (version 12.9.0) to account for the spectral analysis and the estimate of the $0.3-10$ keV band flux.

\subsection{Timing analysis in the $X$-rays}

For each source target of this study, the barycentric and background corrected light curves were extracted (and synchronized) in the soft (0.3-2 keV) and hard (2-10 keV) bands with {several bin of $10$, $60$  and $120$ seconds (see Figure \ref{fig_lc})}.

By using the light curves with a bin size of $10$ seconds and the Lomb-Scargle technique (\citealt{scargle1982}), we blindly performed a search for periodicities in the range between $2 \Delta t$ (being $\Delta t$ the time series bin size) and one third of the observational window. Here we conservatively require to have at least three full cycles for the maximum period tested. This requirement corresponds to a maximum period tested of $\simeq 50$ minutes and $\simeq 167$ minutes for HP Cet and Swift J0820.6-2805, respectively.

\begin{figure*}
  \centering
  \subfloat[]{\includegraphics[width=0.5\textwidth]{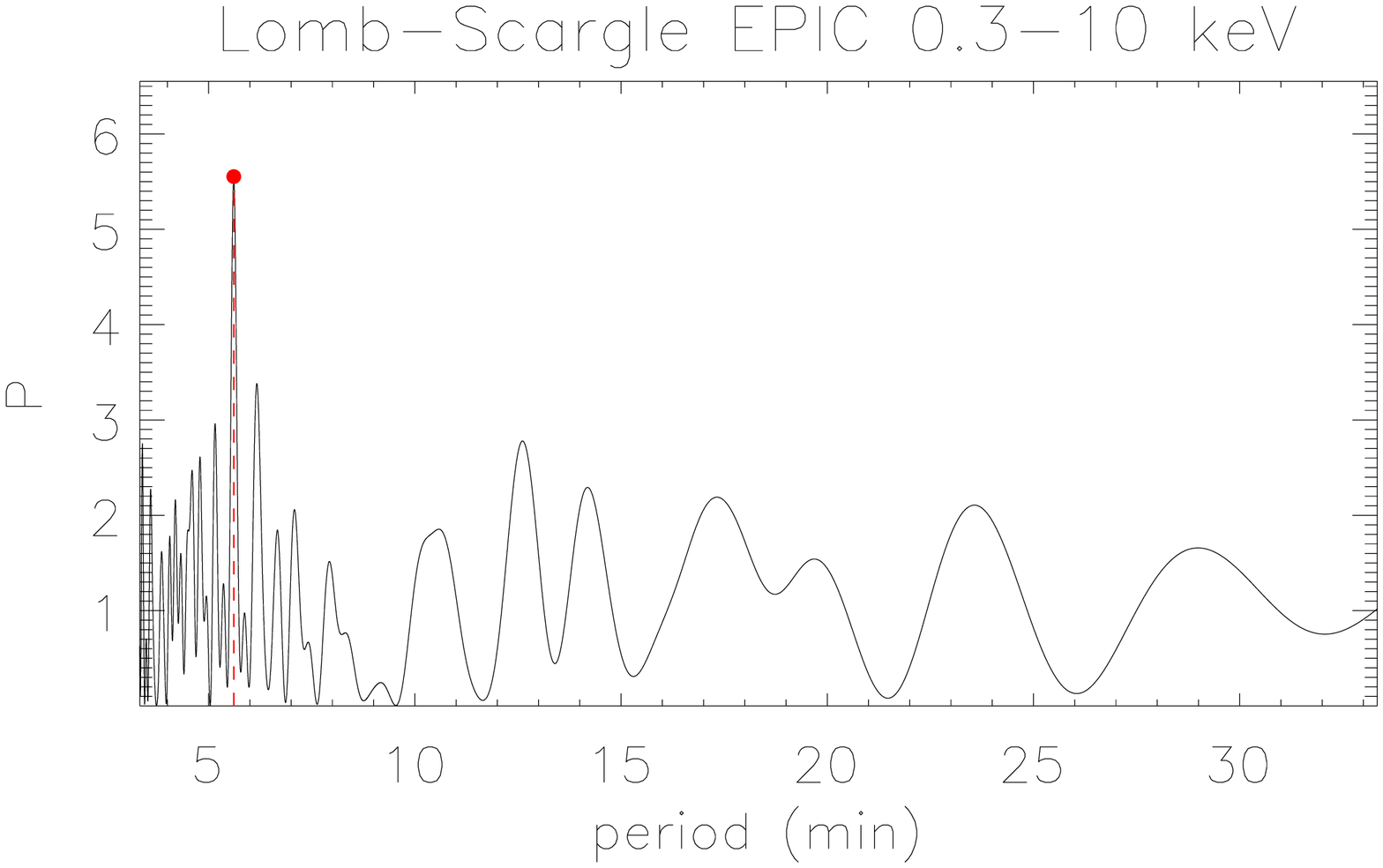}\label{fig3a}}
  \hfill
  \subfloat[]{\includegraphics[width=0.5\textwidth]{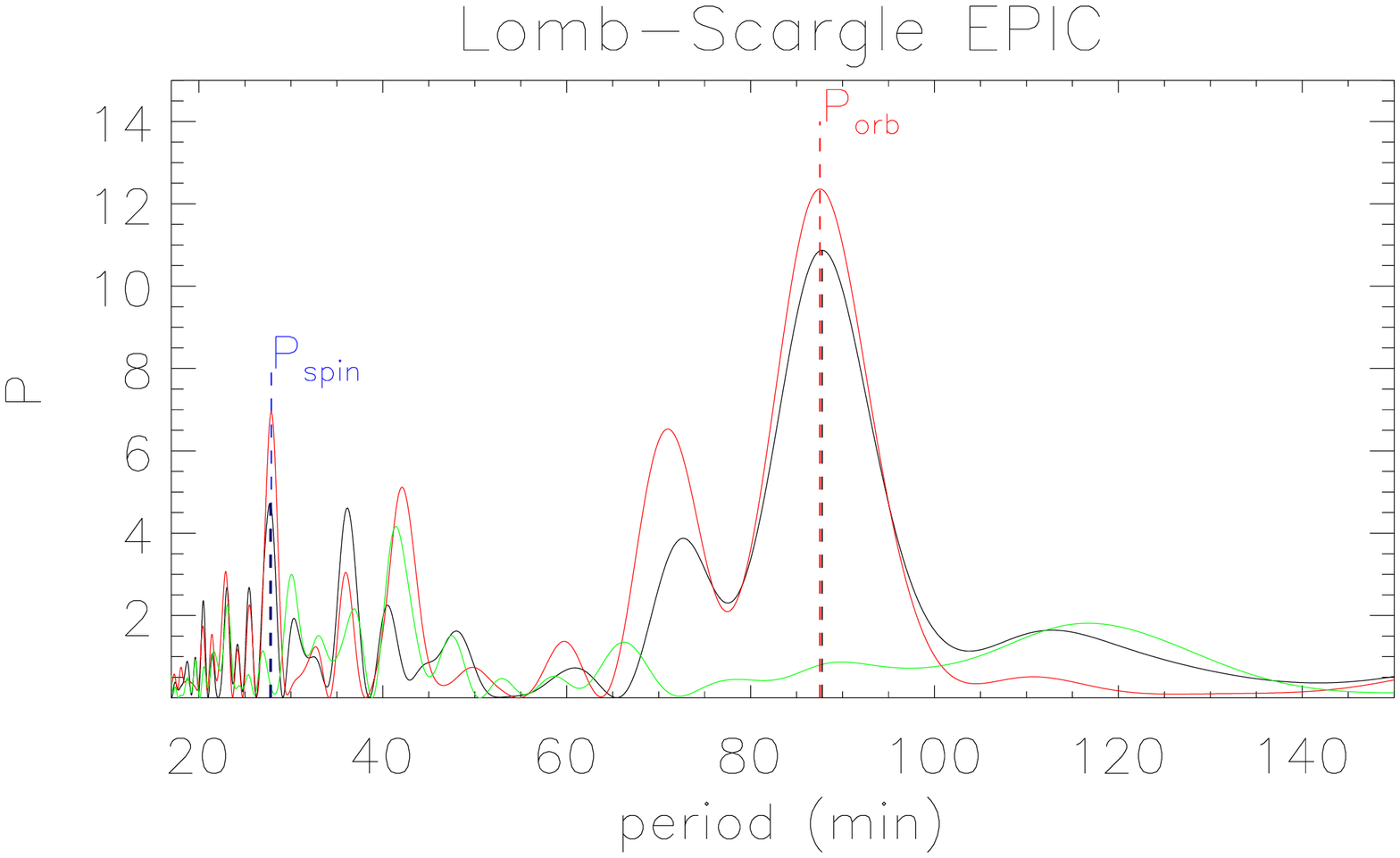}\label{fig3b}}
  \caption{Left panel: the Lomb-Scargle periodogram associated to the HP Cet light curve in the $0.3-10$ energy band ($10$ second bin size). The red vertical line identifies the period found by the method described in the text and corresponds to $\simeq 5.60$ minutes. {In the right panel, we give the Lomb-Scargle periodogram for the soft (red), hard (green), and full (black) $X$-ray light curves. Here, we indicate the orbital period $P_{orb}$ and tentatively identify the WD spin as the period at $\simeq 27.87$ minutes (see text for details).}}
  \label{fig_lombscargle}
\end{figure*}
The result of this analysis is shown in Figure \ref{fig_lombscargle} for both HP Ceti and 
Swift J0820.6-2805. In each panel, we label the interesting periods identified by the method described above. 

The optical light curve of HP Cet (also known as SDSS J0233) shows several modulations with short time scales ranging from $\simeq 60$ minutes up to $160$ minutes  (see, e.g. \cite{southworth2006} and references therein). Having determined spectroscopically a periodic feature at $\simeq 96$ minutes, {these authors tentatively interpret it as the CV orbital period } and the signals at $\simeq 60$ minutes and $\simeq 160$ minutes as the WD spin and the beat between orbital and spin period.  Based on these periods, \citet{southworth2006} classify HP Cet as an IP with a low accretion rate. This is in agreement with the $X$-ray variability observed by \citet{szkody2002} with timescale of $\simeq 500$ seconds as resulting from the analysis of ROSAT data. In this case, the author found that the source is active at the level of $0.0058$ count s$^{-1}$. However, due to the quality of the ROSAT data, it was not possible to get any firm conclusion on the genuineness of the periodicity.

Since $XMM$-Newton data show a $0.3-10$ keV rate  of HP Cet  ($\simeq 0.021$ count s$^{-1}$) larger rate than in the ROSAT case, the variability or periodicity can be searched more easily. As clear from Figure \ref{fig_lombscargle}, panel (a), the Lomb-Scargle periodogram of the HP Cet light curve (in the energy band $0.3-10$ keV) allows us to identify a clear {peak at $\simeq 5.60$ minutes. The data folded at  this period clearly show a repeated structure.}
\begin{figure*}
  \centering
  \subfloat[]{\includegraphics[width=0.5\textwidth]{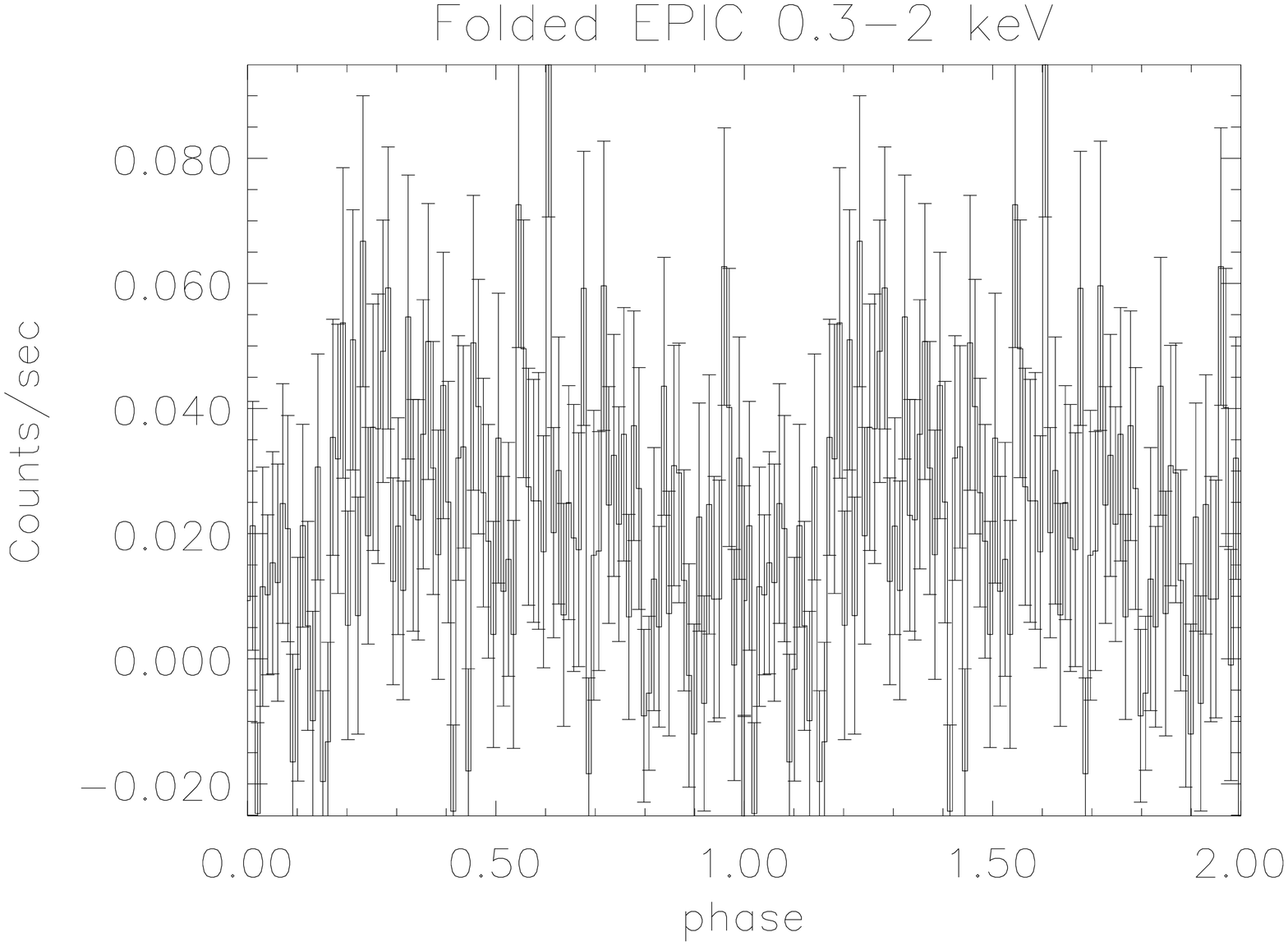}\label{fig4a}}
  \hfill
  \subfloat[]{\includegraphics[width=0.5\textwidth]{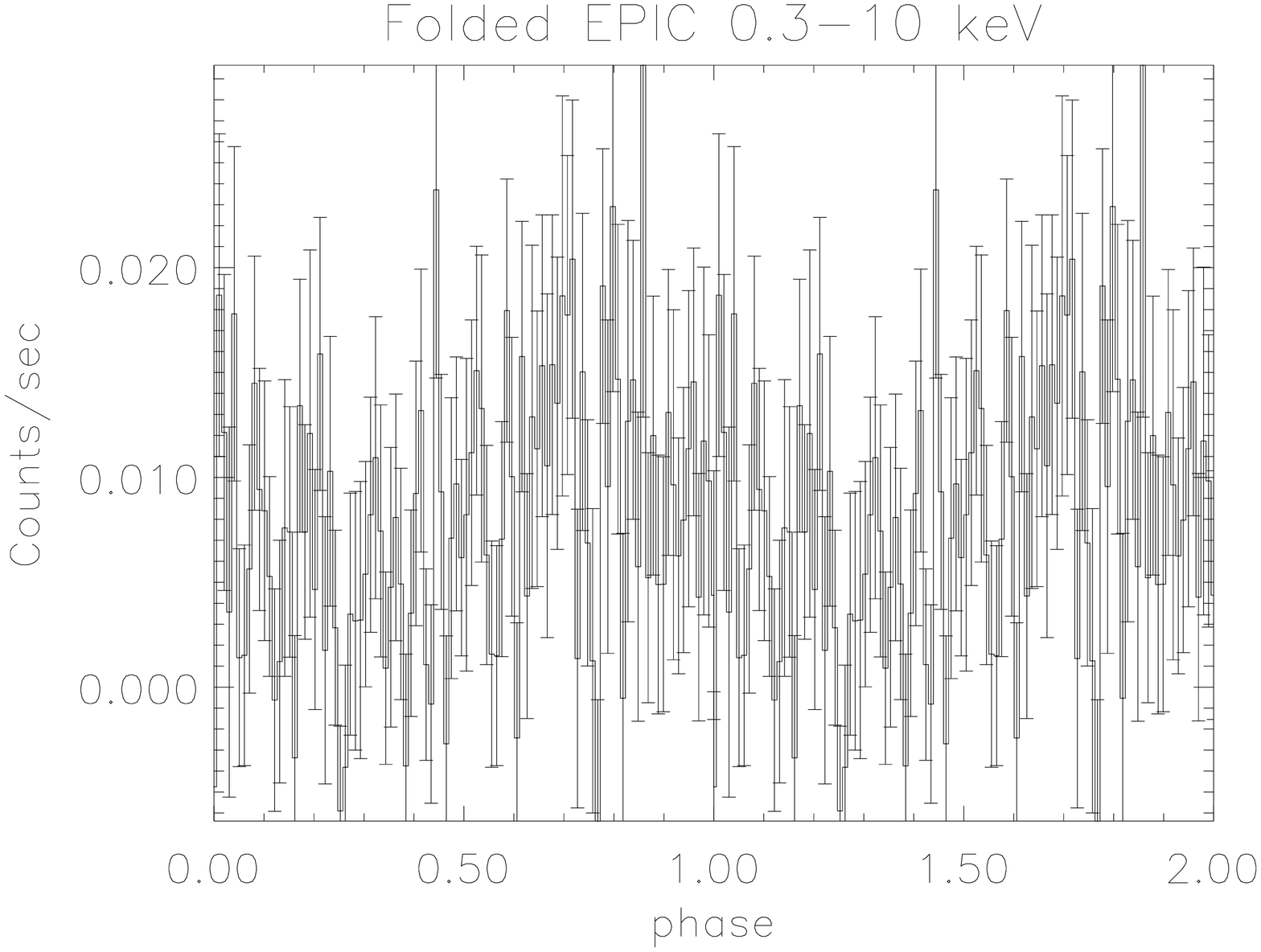}\label{fig4b}}
  \caption{Left panel: the folded light curve (with 60 seconds bins) of HP Cet over the period $\simeq 5.5$ minutes clearly shows a repeated structure. Right panel: here we give the $0.3-10$ keV light curve of Swift J0820.6-2805 when folded at the orbital period $\simeq 87.53$ minutes and with 60 seconds bins. In each panel, the zero phase bin is associate to the first entry in the corresponding light curve (see text for details).}
  \label{fig_folded}
\end{figure*}
{Due to the quality of the data, and to the conservative data reduction procedure described in the previous section,} the $0.3-10$ keV time series of HP Cet do not allow to determine the orbital period of the source ($\simeq 96.08$ minutes ) neither the WD spin (tentatively identified with a periodic feature at $\simeq 60$ minutes) (see, \citealt{southworth2006}) by using the high energy data alone (see in the following for details on the OM data analysis). The HP Cet light curve folded at $\simeq 5.5$ minutes (see Figure \ref{fig_folded}, panel (a)) clearly shows a periodic sinusoidal structure.
We tested the detected periodicity versus spurious features by computing the Lomb-Scargle periodograms of $5\times 10^4$ fake light curves. For each light curve we randomly shuffled the rate values while keeping the stamps of the time axis unchanged. Then, we define the false alarm probability (FAP) at a given tested period as the fraction (w.r.t. the total amount of simulations) of those periodograms having a power larger than that in the power spectrum observed at the same period. Therefore, the probability that a {peak} is genuine can be evaluated as $P=1-FAP$. When we apply this method to the HP Cet periodogram, we found that the $\simeq 5.5$ minutes has a probability of $\simeq 99.5\%$ to be genuine.

{As far as the source Swift J0820.6-2805 is concerned, the $0.3-10$ keV light curve shows an average count rate of $\simeq 0.015$ count s$^{-1}$}. The observational window of the target ($\simeq 35$ ks) allows us to test longer periods by using the Lomb-Scargle technique. {We searched for periodicities in the soft ($0.3-2$ keV), hard ($2-19$ keV), and full ($0.3-10$ keV) light curves independently and the resulting periodograms are given in panel (b)  of Figure  \ref{fig_lombscargle} with red, green and black lines, respectively. 
As in the case of other IPs, Swift J0820.6-2805 shows modulations in the soft band larger than in the hard band.  {$X$-ray signals are often characterized by modulations (\citealt{parker2005}) on the WD spin $P_{spin}$ and the orbital period $P_{orb}$}. In this case, we identify a strong peak, probably associated to the orbital period $P_{orb}\simeq 87.53$ minutes, which almost disappears in the $2-10$ keV band since the source is intrinsically fainter at these energies. The periodogram shows a second peak at $\simeq 27.87$ minutes that we tentatively associate with the spin period of the WD. Note that the power spin peak decreases in the hard band and appears again in the full band where the count rate is still dominated by the $0.3-2$ keV photons. As clearly explained in \citet{warnerbook}, the presence of reprocessing sites and the existence of a modulation on the orbital period often induces the appearance of multiple orbital sidebands. In particular, we expect to find a sideband at  $P_{syn}^{-1}=P_{spin}^{-1} - P_{orb}^{-1}$ which, for the estimated values of spin and orbital periods, corresponds to $P_{syn}\simeq 40.89$ minutes which is close to the peak at $\simeq 41.4$--$42.1$ minutes\footnote{We note that a  period of $\simeq 41.4$ minutes was found by \citet{Halpern1and2015} during one night observation and regarded as the possibe WD spin period.}
 appearing in the periodogram. By considering the full  width at half  maximum as the uncertainty associated to each of the estimated periods (with, in particular, $7$ and $1$ minutes for the orbital and spin periods, respectively), the observed feature appears to be consistent with the synodic period as expected.

We then folded the $0.3-10$ keV light curve at the orbital period of $\simeq 87.53$ minutes and 60 bins per cycle and observed a clear sinusoidal pattern (see Figure \ref{fig_folded}, panel b, where the zero phase is associated to the start for the $XMM$-Newton EPIC data at 55941.9098). 
Finally, having recognized the existence of such features, Swift J0820.6-2805 can be easily classified as a member of the IP class (see also \citealt{nucita2019}, where the same method has been applied to the case of DW Cnc).
}

\subsection{Timing analysis in the UV and B bands}
HP Cet and Swift J0820.6-2805  have been observed by the  Optical Monitor on-board the $XMM$-Newton telescope in imaging and fast mode and by using the UVM2 and B filters. We then binned the light curves in 10 seconds wide bins and used the Lomb-Scargle method to search for periodic features.
\begin{figure*}
  \centering
  \subfloat[]{\includegraphics[width=0.5\textwidth]{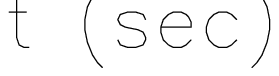}\label{fig5a}}
  \hfill
  \subfloat[]{\includegraphics[width=0.5\textwidth]{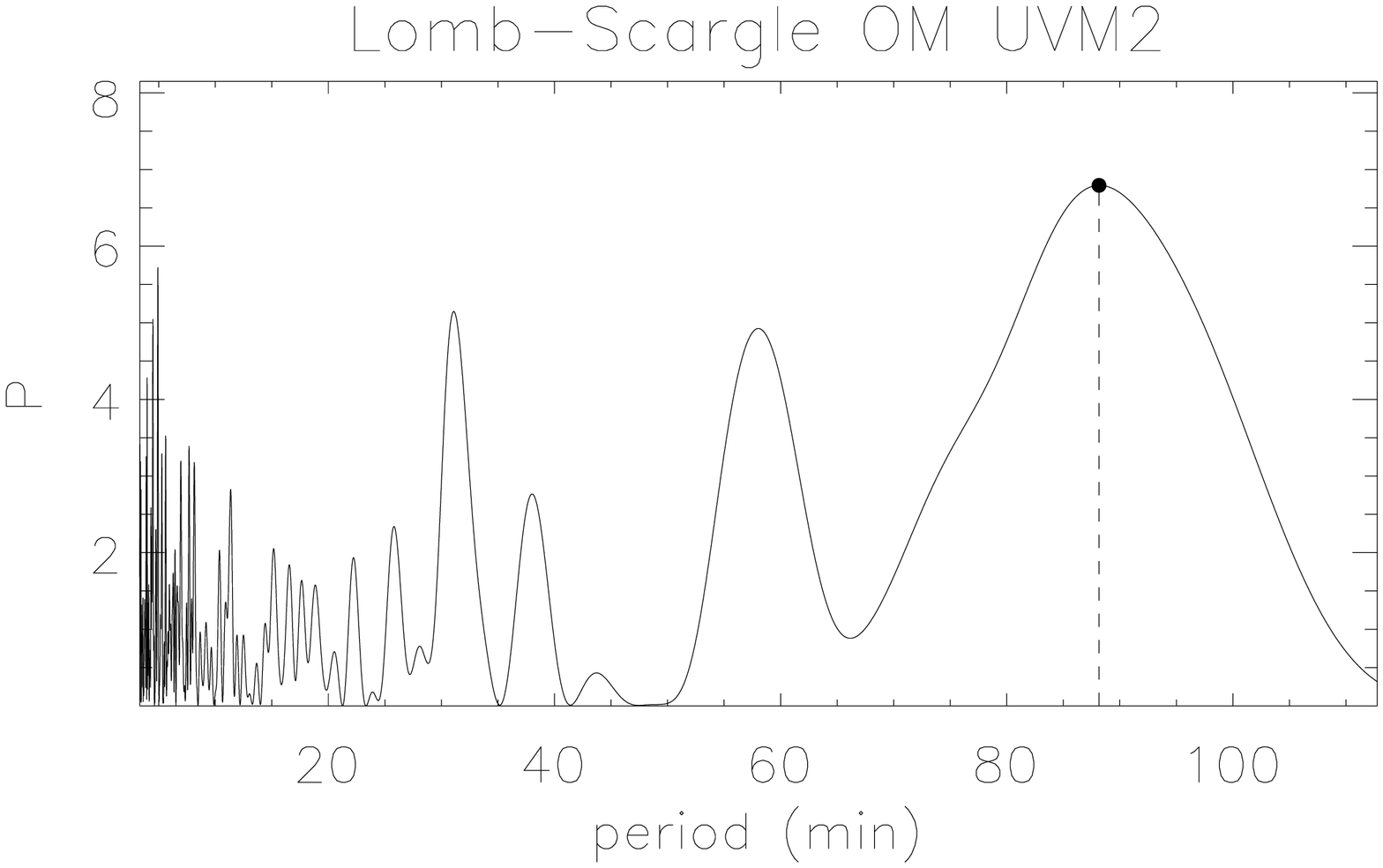}\label{fig5b}}
  \caption{Left panel: the OM light curve (UVM2 filter) of HP Cet with a 60 seconds bin size.In the right panel we give the Lomb-Scargle periodogram with the identification of a $\simeq 88$ minutes period with a confidence level of 1 $\sigma$. The large large {peak} at $\simeq 58$ minutes is very close to the periodicity of approximately $60$ minutes resolved by \citet{southworth2006} and possibly associated to the WD spin period.}
  \label{omlightHPCET}
\end{figure*}

In the case of HP Cet, the useful light curve (Figure \ref{omlightHPCET}, left panel) lasts for $\simeq 5.6$ hours and has an average UVM2 magnitude of $\simeq 19.5$. This allows us to search for any period up to $\simeq 110$ minutes when requiring to have at least three full cycles.  The analysis resulted in the periodogram shown in
Figure \ref{omlightHPCET} (right panel) from which it is clear the presence of a {rather broad peak at $\simeq 88$ minutes
that is marginally consistent with the $\simeq 96$ minutes period found by \citet{southworth2006} analyzing medium-resolution VLT/FORS2 spectroscopy of the source (see their Table 3).} More interestingly, the periodogram shows a large {peak at $\simeq 58$ minutes}, i.e very close to the periodicity of approximately $60$ minutes resolved by \citet{southworth2006} and explained as the the possible WD spin period. In Figure \ref{omfolded} (left panel), we present two cycles of the HP Cet light curve (OM UVM2 filter) folded over the $88$ minutes {and setting the zero phase bin to the starting time of the EPIC $0.3-10$ keV light curve (with 10 seconds bins), i.e. MJD=55941.9098.}

\begin{figure*}
  \centering
  \subfloat[]{\includegraphics[width=0.5\textwidth]{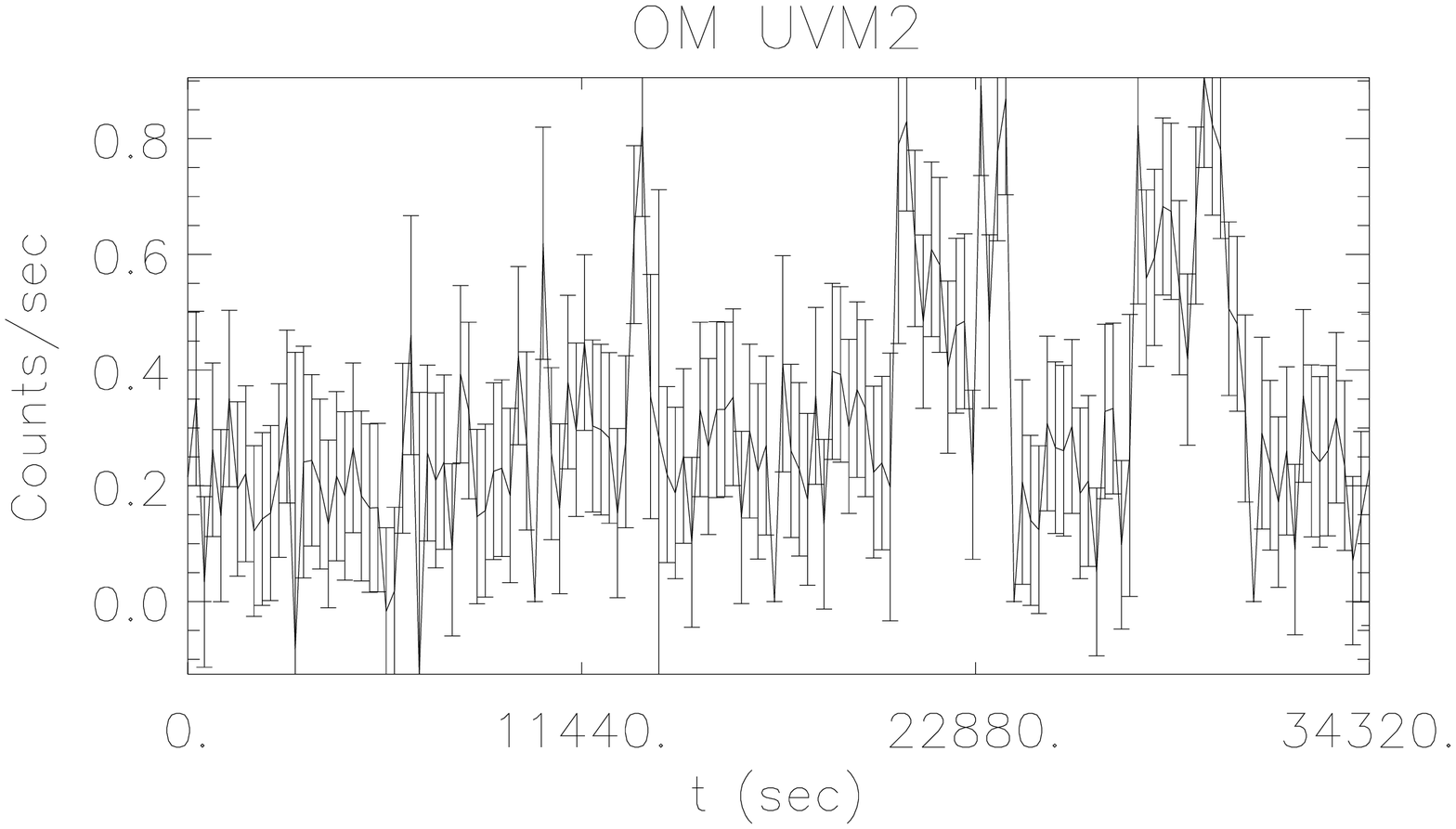}\label{fig6a}}
  \hfill
  \subfloat[]{\includegraphics[width=0.5\textwidth]{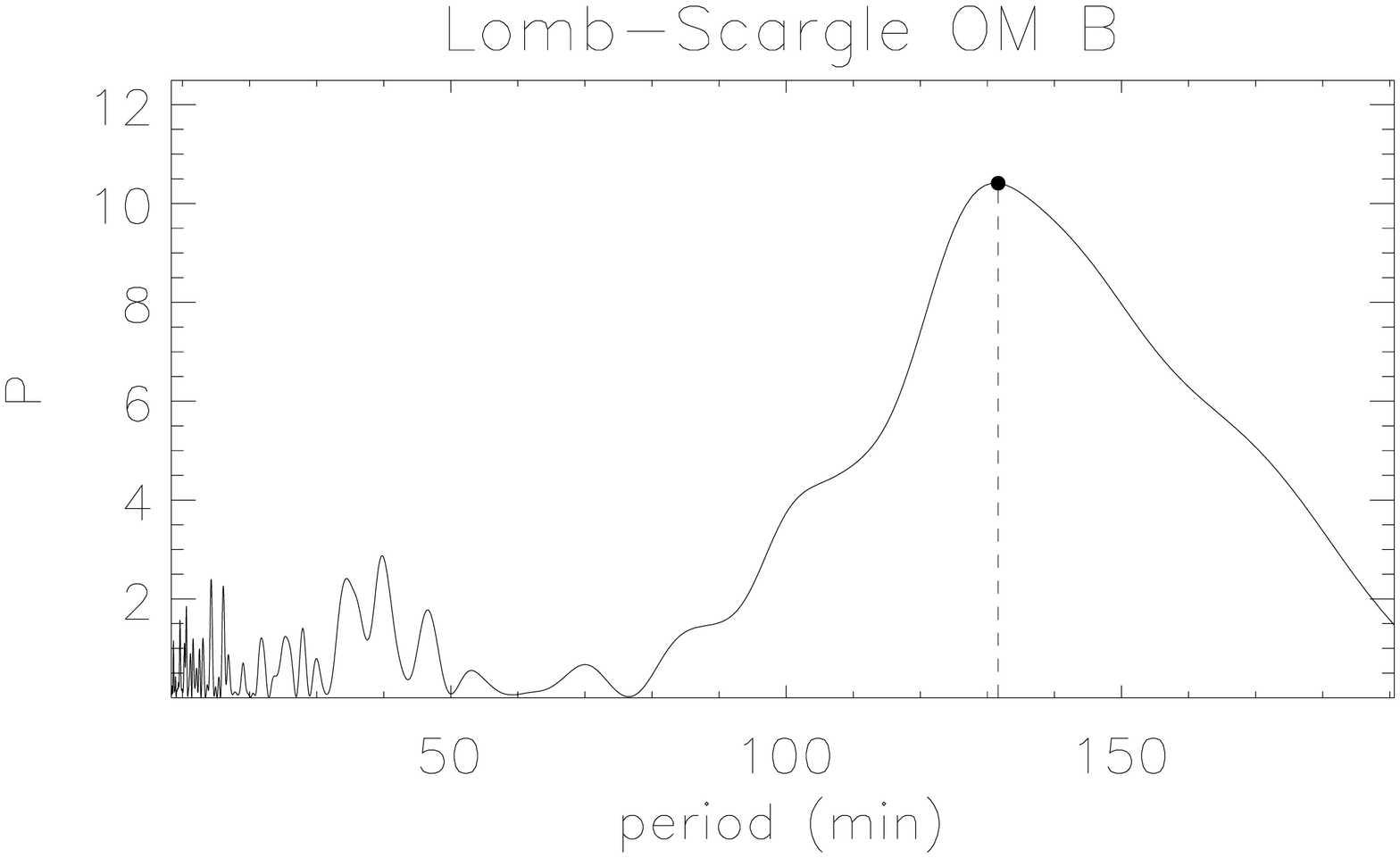}\label{fig6b}}
  \caption{In the left panel we give the OM light curve (B filter) of Swift J0820.6-2805 with a 240 seconds bin size. The right panel shows the Lomb-Scargle periodogram with the identification of a $\simeq 131$ minutes period (see text for details).}
  \label{omlightSWIFT}
\end{figure*}

\begin{figure*}
  \centering
  \subfloat[]{\includegraphics[width=0.5\textwidth]{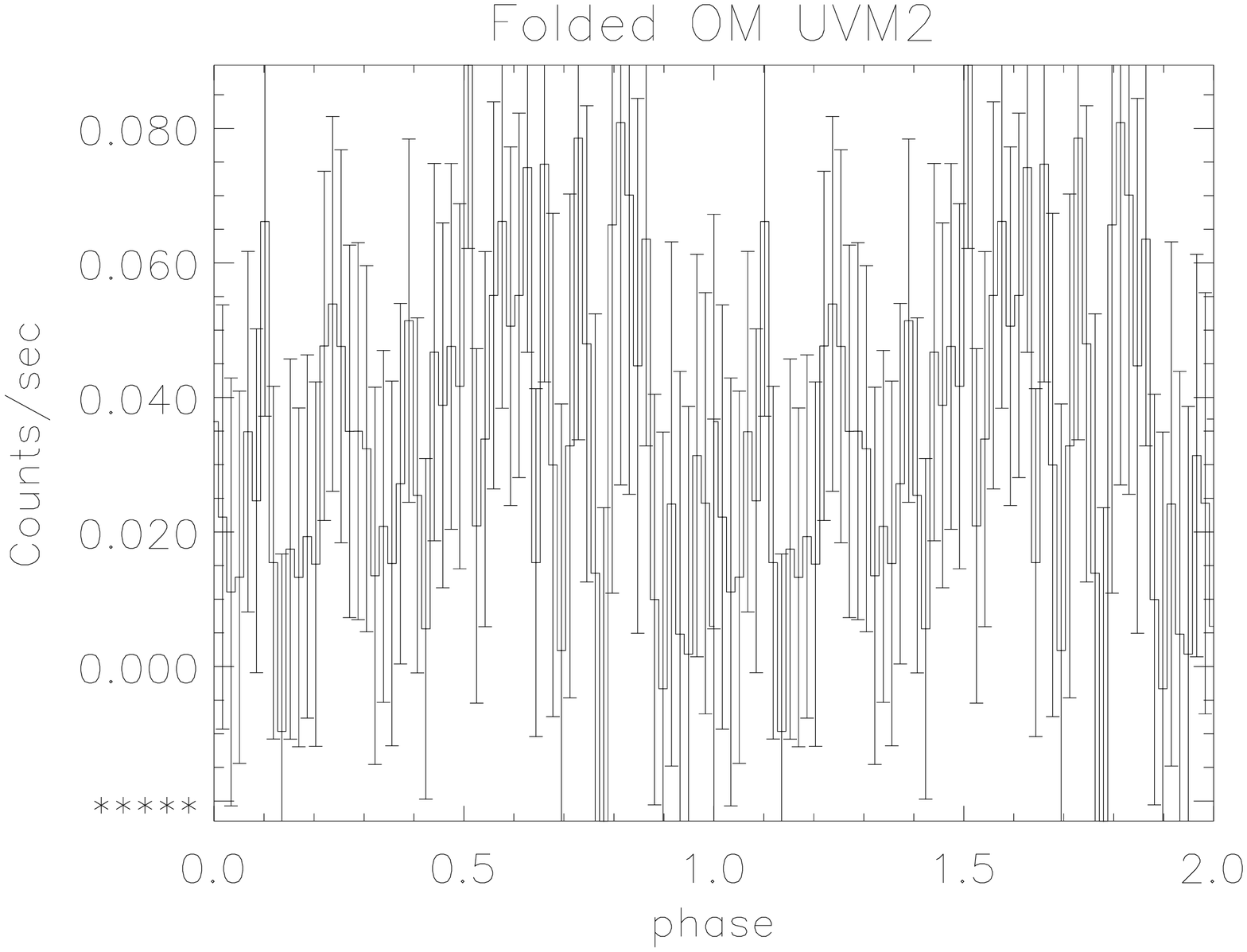}\label{fig7a}}
  \hfill
  \subfloat[]{\includegraphics[width=0.5\textwidth]{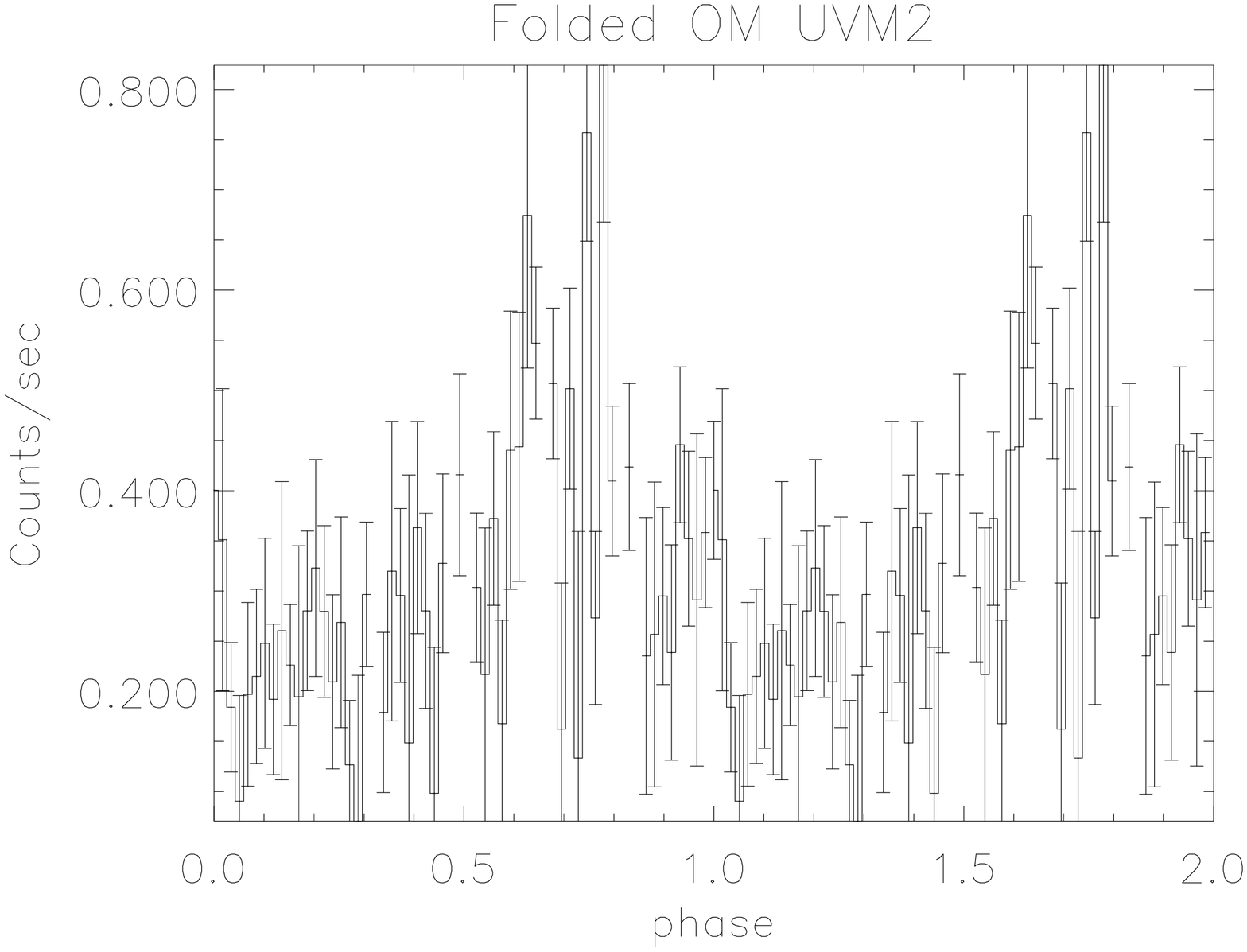}\label{fig7b}}
  \caption{In the left panel we give the OM light curve folded over the 88 minutes of HP Cet. In the right panel the folding of the Swift J0820.6-2805 time series is performed over the variability of $\simeq 130$ minutes detected in the B band.}
  \label{omfolded}
\end{figure*}

{We performed a similar analysis for the source Swift J0820.6-2805 whose light curve (lasting for $\simeq 9.5$ hours) in the B filter and with a bin size of $240$ seconds is given in Figure \ref{omlightSWIFT} (left panel). The average magnitude of Swift J0820.6-2805 is $20.49$. As one can note, the light curve seems to be characterized by a variability on time scales of hundreds of minutes with, in particular, the appearance of two regions of large count rates at $t\simeq 22000$ seconds and $t \simeq 30000$ seconds, respectively. This is also confirmed by the Lomb-Scargle periodogram which shows a rather wide feature at $\simeq 130$ minutes
corresponding to the intrinsic variability of the source on this time-scale.  The existence of such long term variability is also confirmed when folding the light curve at 130 minutes (see right panel in Figure \ref{omfolded}). Here, we set the zero phase bin to the starting time (MJD=58226.3432) of the EPIC $0.3-10$ keV light curve (with 10 second bins).}     

\subsection{Spectral analysis}
{HP Cet and Swift J0820.6-2805 are very faint sources, as it is  clear from the average count rate in the $0.3-10$ keV band of $\simeq 0.02$ counts s$^{-1}$ and $\simeq 0.015$ counts s$^{-1}$, respectively.}
{
For HP Cet, we extracted the spectrum for the source and background regions along with the response matrix files for MOS 1, MOS 2 and pn cameras. We then binned the spectra with the {\it grppha} tool by requiring to have at least 25 counts per energy bin. The spectra were imported within the XSPEC package (version 12.9.0) for the spectral analysis and fitting procedure.
\begin{figure*} 
  \centering
  \includegraphics[width=0.5\textwidth, angle=-90]{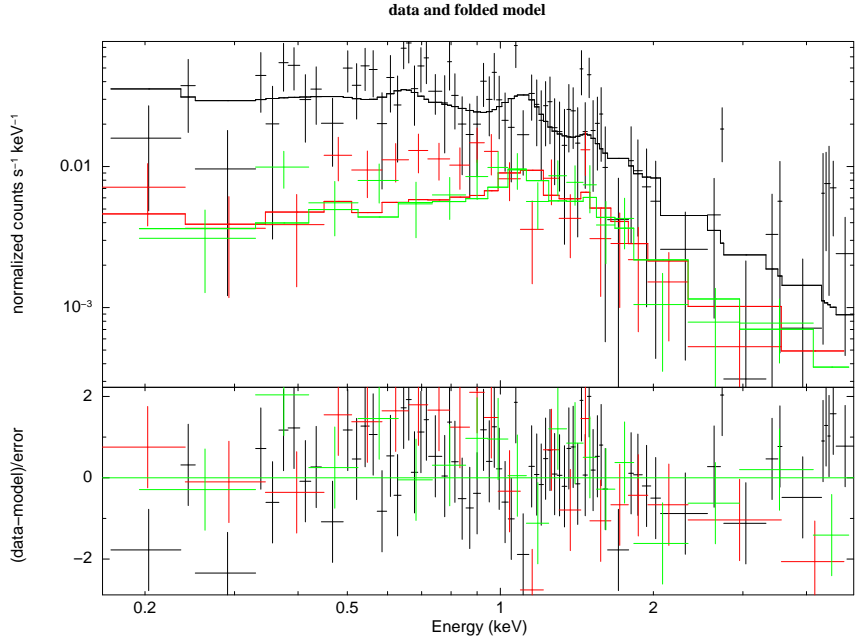}
  \caption{The $0.3-4$ keV spectra for the MOS 1 (red), MOS 2 (green) and pn (black) data of HP Cet together with the best-fit model (see text for details) are presented.}
  \label{spectrumhpcet}
\end{figure*}

Due to the low count rate, the HP Cet spectrum is of poor quality. We first tried to fit it with  a {\it mekal} model describing the emission from a hot diffuse gas \citep{mekal} with the inclusion of emission lines from several elements. 
The estimated distance to HP Cet, as reported in the second data release of the Gaia telescope measurements  (see \citealt{gaiadr2}), is $\simeq 600$ pc, so that we fixed the redshift parameter to zero and the material abundances to the solar values.
We also accounted for any possible absorption by considering a multiplicative component {\it phabs} which depends on the energy as $\exp(-NH\sigma(E))$, where $NH$ is neutral hydrogen column density and $\sigma(E)$ is the photoelectric cross section. The initial value of $N$ was set to the value observed towards the target and provided by \cite{nhtool}, i.e. $NH\simeq 0.0281\times 10^{22}$ cm$^{-2}$. The best fit ($\chi^2=1.1$ for 172 d.o.f.) converged towards a mekal temperature of $kT_{mk}=3.9^{+1.5}_{-0.9}$ and an upper limit on the neutral hydrogen column density of $NH<0.014\times 10^{22}$ cm$^{-2}$. Finally, the unabsorbed $0.3-10$ keV band flux is $(1.3^{+0.2}_{-0.2})\times 10^{-13}$ erg s$^{-1}$ cm$^{-2}$. By using the above distance, the intrinsic luminosity is estimated to be $\simeq 5\times 10^{30}$ erg s$^{-1}$ in the above-mentioned band.
}

{
Although Swift J0820.6-2805 has approximately the same count rate as HP Cet, the longer observation, and the absence of strong solar flares, allowed to collect a larger number of photons which, in turn, enabled us to produce a $0.3-10$ keV spectrum of acceptable quality. We extracted the spectrum of the source, corrected it for the background and exposure and binned the result in order to have at least $25$ counts per energy bin. The resulting spectra for MOS 1 (red), MOS 2 (green) and pn (black) cameras are shown in Figure \ref{spectrumswift}.

By using XSPEC, we started by fitting the data with a single mekal model. As before, since the source is at a distance of  $\simeq 200$ pc (\citealt{parisi2014}), we fixed the redshift parameter to zero and the abundances to the solar values. Note that the data show a clear excess of photons at low energies suggesting that the absorption due to the hydrogen column density $NH$ is likely to be small and that a soft spectral component is required in order to obtain a good fit. In fact, a single mekal component (not absorbed by any intervening matter) was unable to adequately fit the data ($\chi^2$ larger than 2.5) and, consequently,  we added a second thermal component (a {\it bbodyrad} model) to the previous one. 

The best fit procedure (with reduced $\chi^2=1.5$ for 63 d.o.f)  resulted in the mekal and black-body temperatures of $kT_{mk}=3.9^{+1.5}_{-0.9}$ keV and in 
$kT_{bb}=(2.12^{+0.6}_{-0.5})\times 10^{-2}$ keV, respectively. Note that the errors are quoted at the $90\%$ confidence level.  The estimated mekal normalization results to be $N_{mk}=(3.85^{+0.36}_{-0.34})\times 10^{-5}$ in units of $10^{-14}\int n_e n_H dV/4\pi [D_A (1+z)]^2$, where $D_A$ is the angular diameter of the source, $n_e$ and $n_H$ the electron and hydrogen density and the integral is on the emission volume. For the black-body component, the associated normalization $N_{bb}=(1.3^{+89.0}_{-1.3})\times 10^{7}$ is given in units of $R^2_{\rm km}/D^2_{\rm 10~ kpc}$ where $R_{\rm km}$ is the radius (in kilometers) of the $X$-ray emitting region and $D_{\rm 10~ kpc}$ is its distance in units of 10 kpc. Hence, for the assumed distance to the target, the equivalent radius of the source emitting region is in the range $12-600$ km, i.e. much smaller than the radius of the WD itself. We emphasize that \cite{worpel2015} found a quite similar result analyzing the polar CV labeled as  CSS081231:071126+440405).
\begin{figure*} 
  \centering
  \includegraphics[width=0.5\textwidth, angle=-90]{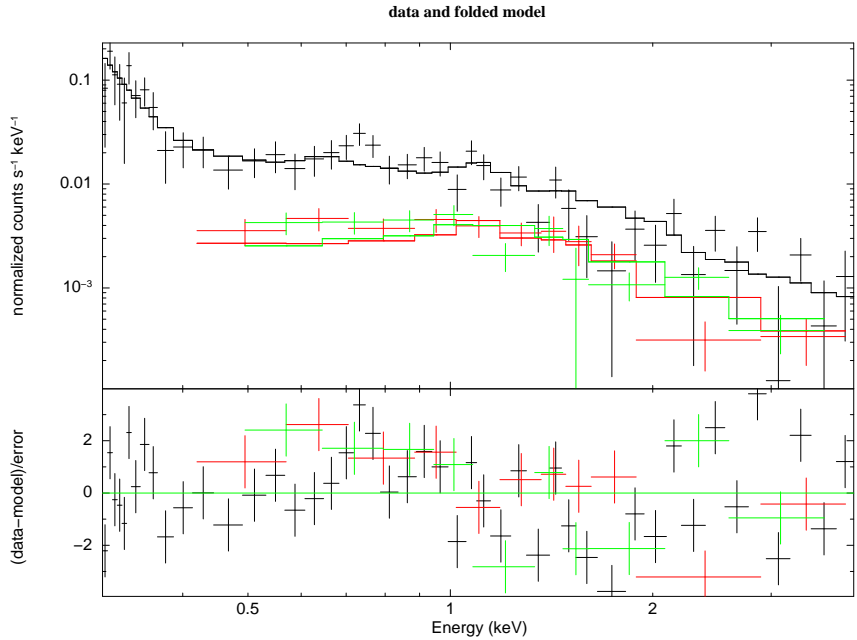}
  \caption{The $0.3-4$ keV spectra for the MOS 1 (red), MOS 2 (green) and pn (black) data of Swift J0820.6-2805 along with the best-fit model (see text for details) are shown.}
  \label{spectrumswift}
\end{figure*}
We then determined the upper limit of the hydrogen column density that absorbs the X-ray flux by introducing a $\it phabs$ component in XSPEC. In particular, we increased the value of $NH$ (e.g., by using the {\it steppar} command in XSPEC) until no statistically acceptable fit could be obtained, i.e. when the $\chi^2$ value changed to $2.7$ with respect to the value obtained by the  best fit. This procedure resulted in a $90\%$  upper limit of $0.0075\times 10^{22}$ cm$^{-2}$, i.e. well below the average galactic column density of $0.293\times 10^{22}$ cm$^{-2}$ towards the target \citep{nhtool}. Below the estimated upper limit, the fit is essentially insensible to the $NH$ value since the absorption is simply not required.

Finally, we evaluated the unabsorbed $0.3-10$ keV band flux of the source to be  $(7.8^{+0.5}_{-2.6})\times 10^{-14}$ erg s$^{-1}$ cm$^{-2}$ corresponding to an intrinsic luminosity of  $\simeq 3.8\times 10^{29}$ erg s$^{-1}$. This appears to be one of the lowest luminosity IPs in X-rays.
}

\section{Discussion and results}
\label{discussion}
{
Intermediate polars are cataclysmic variables, i.e. binary systems in which a strongly magnetized white dwarf (with magnetic field of up to $\simeq 10$ MG) accretes matter from a donor star. The radially infalling material produces a shock wave above the magnetic pole on the surface of the primary star. As a consequence, IPs emit $X$-rays  originating from hot plasma with temperature up to 20-50 keV. Depending on the physical parameters of the accretion gas column, the $X$-ray spectrum can show either cooling flows or photoionization components (see, e.g. \citealt{mukai2003}) which, in addition, might be absorbed by the intervening matter, being the suppression strongly dependent on the actual geometry of the system. In this respect, satellites with large sensitivity (as $XMM$-Newton) allow to discover and study a population of soft IPs characterized, for example, by black-body components with low plasma temperature ($kT\simeq 50$ eV).
 
Altought IPs are expected to be common in the Milky Way, they are difficult to detect because either they  are  intrinsically faint or  they suffer of absorption. Nowadays, it  appears quite clear that faint IPs do exist. In fact, \cite{pretorius} showed that, while most of the detected hard IP sources are characterized by a typical luminosity of $\simeq 10^{33}$ erg s$^{-1}$, there is evidence of the existence of a rather rare faint population. Among the elusive sources (with luminosity less than $\simeq 10^{32}$ erg s$^{-1}$), only a few were observed in the soft (AE Aqr, DQ Her and V902 Mon)  and hard (DO Dra, V1025 Cen and EX Hya) band. Apparently, two unconfirmed IPs (V597 Pup and V475 Sgr, see \citealt{mukai2020}) are at the lower end of this distribution since the source estimated luminosities are of the order of $10^{31}$ erg s$^{-1}$).

As observed by \cite{gansike2005} (but see also \citealt{aungwerojwit2012}), IPs above the period gap (2--3 hr) are characterized by a ratio of $P_{spin}/P_{orb}$ widely distributed in the range $0.01$--$0.1$ while all the systems below the gap have $P_{spin}/P_{orb}>0.1$. Furthermore, since most of the known IP have orbital periods above the gap, it  has been suggested that IPs evolve to polars becoming synchronized (\citealt{can1984}). Of course, if the subclass of low luminosity IP contains short period systems then short period IPs should be intrinsically common as polars. In this case, there is the possibility that long period IP can evolve trough the period gap simply becoming short period IPs.

In this work, we reported on the timing analysis of $X$-ray data from two CVs of interest, HP Cet and Swift J0820.6-2805, which are classified as IP candidate and, as such, listed in IPhome catalogue.  The $XMM$-Newton observations for HP Cet (ID 673140201) and  Swift J0820.6-2805 (ID 0801780401) were not previously analyzed and, in particular, a confirmation of the IP nature for  Swift J0820.6-2805 was missing. Our analysis of the 0.3-10 keV band and UV data of HP Cet confirms the existence of a periodicity of  $\simeq 88$ minutes that we address as the orbital period. Note that this is slightly different from the value of $\simeq 96$ minutes reported by \citet{southworth2006}. We also detected a periodic signal at $\simeq 5.6$ minutes that seems to be a genuine feature. Note that a close variability of about $8$ minutes was found by \citet{szkody2002} analysing past ROSAT satellite data. 

In the case of the IP candidate Swift J0820.6-2805, we showed that the high energy light curve is characterized by a periodic signal on the scale of the orbital period ($\simeq 87.77$ minutes). We clearly identified also the spin period of the WD (at $\simeq 27.87$ minutes) with the typical sidebands expected for magnetic systems, thus allowing us to confirm the IP nature of the source. Moreover, the analysis of the B band light curve showed also the existence of a $\simeq 130$ minutes variability that needs to be confirmed by further dedicated follow up observation.

Both HP Cet and Swift J0820.6-280 appear to be under-luminous sources in $X$-ray since, for the estimated distances of 600 pc and 200 pc, they have luminosities of
$\simeq 5\times 10^{30}$ erg s$^{-1}$ and $\simeq 3.8\times 10^{29}$ erg s$^{-1}$, respectively. The low luminosity appear to be due to the intrinsic accretion mechanism since the targets show a negligible absorption. The two IPs in question are particularly interesting not only because they are characterized by a very low luminosity but also because their orbital period appears to be well below the so-called period gap. These IPs appear to be members of a currently poorly populated   class (which, in reality, could be the most numerous) formed by short period and low luminosity IPs which probably evolved  from the long period stage to the short period one.  Confirming this scenario requires the discovery and study of other objects of this class, and constitutes a challenging research subject. 
}

\section*{Acknowledgements}
 This  paper  is  based on observations from
XMM-Newton, an ESA science mission with instruments
and  contributions  directly  funded  by ESA Member  States  and  NASA. We thank for partial support the INFN projects TAsP and EUCLID. {We also acknowledge the anonymous Referee for several suggestions which allowed to improve the paper and for spotting a crucial error in the performed analysis.
}

\section{Data availability}
The data underlying this article were accessed from the XMM-Newton Science Archive (\url{http://nxsa.esac.esa.int/nxsa-web/#home}, ID 673140201 and  0801780401). The derived data generated in this research will be shared on reasonable request to the corresponding author.





\bsp	
\label{lastpage}
\end{document}